\documentclass[journal=aesccq,manuscript=article, layout=twocolumn]{achemso}

\usepackage[utf8]{inputenc}
\usepackage{lipsum}
\usepackage[version=4]{mhchem}
\usepackage[T1]{fontenc} 
\usepackage{siunitx}
\usepackage{caption}
\usepackage{subcaption}
\usepackage{comment}
\usepackage{dblfloatfix}
\DeclareSIUnit{\Rydberg}{Ry}
\DeclareSIUnit{\Bohr}{Bohr}
\usepackage{hyperref}   
\usepackage[nameinlink,noabbrev,capitalise]{cleveref}
\usepackage{cleveref}   
\usepackage{xcolor}
\usepackage{contour}
\contourlength{0.1pt}

\author{Anant Vaishnav}
\affiliation[Aarhus University]
{Center for Interstellar Catalysis, Department of Physics and Astronomy, Aarhus University, Aarhus C 8000, Denmark}
\author{Niels M. Mikkelsen}
\affiliation[Aarhus University]
{Center for Interstellar Catalysis, Department of Physics and Astronomy, Aarhus University, Aarhus C 8000, Denmark}
\author{Mie Andersen}
\email{mie@phys.au.dk}
\affiliation[Aarhus University]
{Center for Interstellar Catalysis, Department of Physics and Astronomy, Aarhus University, Aarhus C 8000, Denmark}

\title[Machine learning exploration of binding energy distributions of H$_2$O at astrochemically relevant dust grain surfaces]
  {Machine learning exploration of binding energy distributions of H$_2$O at astrochemically relevant dust grain surfaces}


\begin{document}

\begin{abstract}
Binding energies (BEs) of adsorbates on interstellar dust grains critically control adsorption, desorption, diffusion, and surface reactivity, and therefore strongly influence astrochemical models of star- and planet-forming regions. While recent computational studies increasingly report full distributions of BEs rather than single representative values, these distributions are typically derived for either bare grain surfaces or thick water-ice mantles. In this work, we bridge these regimes by systematically investigating the BE distributions of water on partially and fully ice-covered dust grain surfaces. 

We employ machine-learning interatomic potentials (MLIPs) based on graph neural networks to model water adsorption on graphene and on the Mg-terminated (010) surface of forsterite, representing carbonaceous and silicate grains, respectively. The models enable extensive sampling of adsorption sites on water clusters, monolayers, and bilayers generated under both crystalline (thermally processed) and amorphous (low-temperature) growth conditions. At submonolayer coverage, the chemical nature of the underlying grain strongly affects both ice morphology and binding energies, with Mg--O interactions on silicate surfaces producing particularly deep binding sites. From monolayer coverage onward, adsorption on both substrates is dominated by hydrogen bonding within the ice, reducing the influence of the grain material. 

Across all coverages, amorphous ice structures systematically shift the BE distributions toward stronger binding compared to crystalline ice, introducing highly stable defect and pocket sites. These results demonstrate that BE distributions in the submonolayer to few-layer ice regime are broad and highly surface dependent, and they provide physically motivated input for next-generation astrochemical models incorporating surface heterogeneity.
\end{abstract}

\section{Introduction}

The chemical composition of star-forming regions plays a critical role in their evolution. This composition is primarily governed by transition processes between the gas phase and solid phase, such as adsorption and desorption of atoms or molecules from icy dust grains \cite{Minissale2022}. Explaining astrophysical observations thus requires a deep understanding of gas--grain transitions and the chemical reactions adsorbed species can undergo on grains.
All of these processes are heavily dependent on the BE of the adsorbates, hence, determining BE values allows for better interpretations of observations by including these values in predictive astrochemical models.

Among the most relevant BE values are those related to the binding of adsorbates on dust grain surfaces (primarily carbonaceous and silicaceous materials) and amorphous solid water (ASW), the main constituent of interstellar ices \cite{Boogert2015,McClure2023}. 
Depending on the temperature of the astrophysical environment -- from diffuse and dense clouds to protostellar envelopes and planet-forming disks -- dust grains may expose either the clean surface or be partially or completely covered in ice. Recent laboratory experiments have also suggested that grains and ice may be mixed \cite{Potapov2021} and that the porosity of dust grains means that they may be covered by a submonolayer or few monolayers of ice only, which would make the role of the clean dust grain for surface processes much more important than thick ice mantles.\cite{Potapov2020} For silicate grains specifically, recent laboratory experiments have furthermore suggested that water ice can become trapped in the grain up to temperatures of about 470 K,\cite{Potapov_2024} which has implications for the existence of water in the diffuse interstellar medium and our understanding of the role of water in planet formation.

Hence, water is highly relevant for BE studies, both as part of the grain surface and as the adsorbate. Experimental studies based on temperature programmed desorption (TPD) have provided BEs of water at various astrochemically relevant substrates such as ASW \cite{Fraser2001,Brown2006,Rosu-Finsen2022} and graphite.\cite{Bolina2005}
Quantum chemistry calculations of water BEs have mainly considered cluster models of ASW.\cite{Ferrero_2020,Duflot2021,Bovolenta_2022,Tinacci2023} These studies applied methods ranging from density functional theory (DFT) to QM/MM (Quantum Mechanics/Molecular Mechanics) or high-QM/low-QM approaches such as the ONIOM method, where only a smaller part of the system around the adsorption site is described using the higher level of theory (DFT or coupled cluster).
For carbonaceous and silicaceous materials, water BE calculations have been carried out using DFT and periodic supercells.\cite{Brandenburg2019,Asaduzzaman2013,Molpeceres2018}
In addition to periodic, crystalline silicate surfaces, amorphous nanosilicate clusters are also highly relevant dust grains candidates that have been considered in DFT studies of the energy profile for the formation and binding of molecular water.\cite{Goumans2011}
Finally, machine learning interatomic potentials have recently emerged as a powerful tool for BE studies in astrochemistry, enabling efficient sampling of binding sites on complex substrates such as ASW.\cite{Molpeceres2020,Bovolenta2025}

Astrochemical models -- whether based on kinetic Monte Carlo (KMC) or rate equations -- often rely on BEs for estimating rate constants for both desorption and diffusion processes. For diffusion, it is then assumed that the barrier can be expressed as a fixed fraction of the desorption barrier.\cite{Cuppen_2007,Garrod_2013,Holdship_2017} Traditionally, most models have employed a single BE value for a given molecule on a particular surface (e.g.\ ASW), where the value can come from experimental data or be estimated using simplified schemes such as the linear addition of fragment binding energies. With increasingly detailed computational studies now providing not only average values but full distributions of BEs, the question naturally arises how to incorporate this information in astrochemical models. It is in principle straightforward to do in KMC simulations, which model the spatial distribution of possible binding sites and keep track of the position of every adsorbate. The incorporation of BE distributions in coarse-grained rate equation-based models requires more consideration, but has also recently been addressed.\cite{Furuya_2024}

In the present work, we go beyond the separate consideration of dust grain versus ice surface for BEs. We consider different computational models of partially or completely water ice-covered grain surfaces, where a graphene sheet and the (010) surface facet of forsterite are used as model grain surfaces. MLIPs based on graph neural networks are trained and used both to generate structures of ice-covered grains and to sample BE distributions of water on the resulting surfaces. The distributions are analysed based on the bonding environments of water. We show how the chemical nature of the dust grain strongly influences the structure of the grown ice and the water BE distribution. Furthermore, we quantify the increase in BE caused by amorphous ice structures formed at low temperatures, as compared to the crystalline case. The insights obtained here for water adsorption may be transferable to other astrochemically relevant adsorbates with hydrogen bonding capacity. Hence, our results may form the basis for future studies targeting diffusion or reactivity of cosmic dust grains in the submonolayer to few-layer ice coverage regime.

\section{Methods}

\subsection{Overview}

To cover a range of different astrochemically relevant binding environments, we developed MLIPs both for a water ice-covered silicate (forsterite) surface and a graphene surface. The graphene model was also trained to include \ce{CO2} as an adsorbate. The present work deals however only with \ce{H2O}, and the roles of \ce{CO2} and mixed \ce{H2O}/\ce{CO2} ices will be the topics of future works. To ensure consistent quality of \ce{H2O} BE predictions across the models, a benchmark involving a \ce{H2O} adsorbate on an amorphous \ce{H2O} ice was performed. The results (see Figure S\num{1} in the Supporting Information) show that all the models are able to predict the BE with near-DFT accuracy. The models differ slightly in choice of architecture hyperparameters and settings for the training data generation, and as such their exact details will be described in the following sections.

\subsection{DFT settings}
The training data for the MLIPs were obtained using density functional theory (DFT), the Perdew-Burke-Ernzerhof GGA exchange-correlation functional \cite{Perdew_1996} and plane-wave basis sets.
Calculations for the graphene model were handled with the atomic simulation environment (ASE) package \cite{Hjorth_Larsen_2017} and the GPAW code \cite{Mortensen_2005, Enkovaara_2010}. 

For \ce{H2O}/\ce{CO2} structure searches and molecular dynamics (MD) simulations described in detail in Section \ref{sec:data_acq} below, a plane-wave cutoff energy of \qty{500}{\eV} was used. The graphene lattice constant obtained with these settings was 2.465 Å. The structures selected for use as training data for the MLIP training were recalculated with a \qty{600}{\eV} energy cutoff. The self-consistency cycle was deemed converged when the change in electron density was less than $10^{-5}$ electrons per valence electron, and the maximum change in eigenstates was less than $10^{-7}$ eV$^2$ per valence electron. A Fermi-Dirac distribution with a width of \qty{0.1}{\eV} was used to set orbital occupancies. The Brillouin zone was sampled using $4\times4$ Monkhorst-Pack sampling for $5\times5$ cells, and $8\times 8$ for $3\times3$ cells\cite{Monkhorst_1976}. A D4 correction term \cite{Caldeweyher_2017,Caldeweyher_2019,Caldeweyher_2020} was added to the DFT calculations to account for dispersion interactions.

Calculations for the silicate model were performed with the Quantum Espresso (QE) distribution \cite{Giannozzi_2009,Giannozzi_2017}, with individual atom pseudopotentials as available in the standard solid state pseudopotential library (SSSP) \cite{prandini2018precision}.
The optimization of bulk forsterite (\ce{Mg2SiO4}) was carried out using the Broyden-Fletcher-Goldfarb-Shanno (BFGS) algorithm with a convergence threshold of \qty{2.8e-4}{\Rydberg} (\qty{3.8e-3}{\eV}) for energies and \qty{e-3}{\Rydberg \per \Bohr} (\qty{13.6e-3}{\eV \per \angstrom}) for forces. The plane-wave cutoff used for bulk structures was \qty{50}{\Rydberg} (\qty{680}{\eV}) and the self consistent convergence threshold for energy was set to \qty{e-7}{\Rydberg} (\qty{13.6e-7}{\eV}). Brillouin zone sampling of the bulk unit cell was performed with $8\times8\times8$ k-point grid. The obtained lattice parameters of $a = 4.798, b = 10.313, c = 6.041$ are consistent with experimental values \cite{Kirfel2005} and other theoretical works \cite{geng2019}.

In this work, we concentrate on the nonpolar (010) surface facet with Mg termination, as it has the lowest surface energy among the predominant terminations of forsterite.\cite{deleeuw2000} The surface was obtained by cleaving the bulk structure and used for surface adsorption. All calculations involving the forsterite surface were done with a $2\times2$ k-point grid for $1\times1$ slabs and only the Gamma point for larger slabs, \qty{50}{\Rydberg} (\qty{680}{\eV}) plane-wave cutoff, self consistent energy convergence cutoff of \qty{e-8}{\Rydberg} (\qty{13.6e-8}{\eV}), a D3 correction term \cite{Grimme-D3} to account for dispersion interactions, and a dipole correction  \cite{Dipole-correction-1999}, as implemented in QE. 

\subsection{PaiNN architecture and model settings}

We use the message-passing neural network PaiNN \cite{pmlr-v139-schutt21a} as our model architecture. PaiNN uses a graph representation for input structures, with edges between nodes (atoms) constructed based on a cutoff distance. Node features are updated through a sequence of interaction layers, aggregating information from their immediate neighbours within each layer. The final node features are then used as an input to a neural network with atomwise output layers, predicting properties of interest, such as energies or forces. The PaiNN architecture has previously been shown to accurately model the potential energy surface of various atomistic systems \cite{Kube_ka_2024, Winter_2023}.

The graphene PaiNN model used a \qty{6}{\angstrom} cutoff distance, \num{512} atom basis features, \num{50} radial basis functions and \num{4} interactions blocks, while the silicate model used a cutoff of \qty{5}{\angstrom}, \num{128} atom basis features, \num{20} radial basis functions and \num{3} interactions blocks.

All models were trained on both energies and forces using the AdamW optimizer \cite{adamw} with loss weights of \num{0.05} for energies and \num{0.95} for forces. Training was done with a learning rate scheduler with an initial learning rate of \num{e-4}, which was halved if the validation loss did not improve after \num{20} epochs, together with early stopping if the validation loss stagnated. For the graphene model, a training, validation and testing data split of $80/10/10$ was used, while the silicate model used a $70/20/10$ split. The performance of the models for predicting energies and force components of the test sets are shown in Figure S\num{2} and Figure S\num{3} in the Supporting Information.

To benchmark the performance of the trained graphene PaiNN model, we focus on high-level electronic structure results from the literature,\cite{Brandenburg2019} since the experimental adsorption energy of a single \ce{H2O} molecule on graphene is unavailable due to rapid cluster formation.\cite{Ma2011} As shown in Table S1 in the Supporting Information, the final PaiNN model predicts an  adsorption energy of $-77$~meV, differing only by 13~meV from the diffusion Monte Carlo reference value of $-90$~meV.
For the adsorption of a single \ce{H2O} on forsterite, we benchmark the silicate PaiNN model against the results for an adsorption structure investigated with the hybrid B3LYP functional corrected with D2 dispersion interactions.\cite{Molpeceres2018} As shown in Table S2 in the Supporting Information, the adsorption energy predicted by the final PaiNN model ($-1.14$~eV) differs by only 30~meV from the literature result ($-1.17$~eV).

\subsection{Data acquisition}\label{sec:data_acq}

The training-validation-test data sets for the models were generated using global structure optimization and MD, involving \ce{H2O} and \ce{CO2} on graphene for the graphene model and \ce{H2O} on forsterite for the silicate model. This resulted in diverse datasets with both unstable and highly stable configurations from the beginning and end point of the structure searches, respectively, together with high-temperature MD structures to get non-equilibrium bond lengths. Examples of structures in the data sets are shown in Figure S\num{4} and Figure S\num{5} in the Supporting Information.

Global structure optimization was performed using the GOFEE algorithm \cite{Bisbo_2020, Bisbo_2022} with up to \num{12} molecules for searches on graphene and up to \num{65} molecules for searches on fosterite. Searches on graphene were performed on $3\times3$ and $5\times5$ graphene surfaces, while searches with forsterite present were done with $1\times 1$ and $2\times2$ forsterite slabs consisting of two layers of \ce{Mg2SiO4}. The structures selected for model training were recalculated with a third \ce{Mg2SiO4} layer added to make the data consistent with other data acquisition methods. GOFEE will occasionally produce unphysical structures at the beginning of a run, due to the surrogate model used in GOFEE not yet having sufficient data to describe the underlying potential energy landscape. To avoid adding these structures to the final databases, structures generated for the graphene model with an atomic force above \qty{8}{\eV\per\angstrom} were discarded immediately, while for the silicate model they were discarded after an iterative retraining process described further below.
The final data points from the structure searches included in the graphene data set were selected using farthest point sampling, using the SOAP descriptor \cite{Bart_k_2013} as the target.

The \textit{ab initio} MD simulations for the graphene model were performed at different mixing ratios up to \num{6} molecules at \qty{250}{\kelvin} for \qty{5}{\pico\second}, using the Berendsen thermostat \cite{Berendsen_1984} in the NVT ensemble. All simulations were done on a $3\times 3$ graphene surface. 
Forsterite simulations were done with \num{1}, \num{3} and \num{5} \ce{H2O} on a $1\times1$ $(010)$ forsterite slab at \qty{100}{\kelvin}, using a Nosé-Hoover chain thermostat \cite{Martyna_1992} in the NVT ensemble. 
All \textit{ab initio} MD simulations used a timestep of \qty{0.5}{\femto\second} with a thermostat timescale of \qty{500}{\femto\second}.

The initial silicate model was trained on 5381 structures, consisting of farthest point sampled structures from AIMD runs and all the structures from GOFEE searches. For generating further data, a \qty{100}{\kelvin} MD simulation with 8 \ce{H2O} molecules on the $1\times1$ $(010)$ forsterite slab was performed with the initial model. During the MD run, the model generated unphysical results such as the extraction of Mg atoms from the surface.
To correct this issue, 1172 DFT single-point evaluations of structures from the MD trajectory (with and without Mg extracted) were added, giving a total dataset of 6553 structures.
Next, structures containing forces on any atom greater than \qty{10}{\eV\per\angstrom} in any direction were removed from the database. Then, with this new dataset consisting of 5534 structures, a new PaiNN model was retrained. With this model, no unphysical extraction of Mg atoms was observed in MD.

The stability of the resulting model was further tested by running MD at higher temperatures (\qty{500}{\kelvin}) with 10, 15 and 20 \ce{H2O} molecules. Dissociation of water molecules was observed at the silicate-water interface for all the runs. This is known from previous DFT studies to be unfavorable on the (010) surface facet of forsterite.\cite{Molpeceres2018} 
To correct this issue, 179 DFT single-point evaluations of structures from the MD trajectories were added, giving a total dataset of 5713 structures. The final PaiNN model was trained on this dataset. With this model, no unphysical water dissociation was observed in MD.

\subsection{Generation of surfaces}
\label{subsec:Generation of surfaces}
The final trained PaiNN models were used in the surface generation, where we aim to generate realistic models of interstellar ices on carbon or silicate grains. The morphology of the ice is dependent on its thermal history. We therefore compared two different approaches for structure generation: low-temperature MD for amorphous structures and global structure optimization for more stable and crystalline structures.
To examine different astrochemically relevant binding configurations, such as an adsorbate on bare graphene next to an ice or an adsorbate on top of an ice away from an exposed grain, structures with different grain coverages were generated. We hereby consider three distinct types of structures with different coverages: clusters (where a compact ice leaves most of the underlying grain bare), monolayers and bilayers (where in the latter two cases almost all of the grain is covered).

Structure generation using MD was done in the NVT ensemble with a Nosé-Hoover chain thermostat \cite{Martyna_1992} at \qty{10}{\kelvin}, a timestep of \qty{0.5}{\femto\second} and a total simulation time of \qty{50}{\pico\second}. A chain length of \num{3} and a thermostat time constant of \qty{200}{\femto\second} was used. All the water molecules were placed in the cell before starting the MD simulation. 
For the silicate surface, the lateral diffusion of water is very slow. Hence, to form a compact cluster, we restricted the placement area to within one quadrant of the supercell. This restriction was not necessary for the graphene surface, where the lateral diffusion of water is very fast, even at \qty{10}{\kelvin}.
In all cases, the molecules were placed with a minimal height from the surface of about \qty{2}{\angstrom} and a maximal height of \qty{4}{\angstrom} to \qty{6}{\angstrom}, depending on the desired coverage. The placement also respected a minimal distance of about \qty{2}{\angstrom} between any atom in the molecule to be placed and any other atom in the ice-covered surface. 

For the silicate, the generated surfaces contained 14 \ce{H2O} molecules in a $5\times5$ cell for cluster, 10 \ce{H2O} molecules in a $2\times2$ cell for monolayer and 20 \ce{H2O} molecules in a $2\times2$  cell for bilayer. For the graphene, the generated surfaces contained 35 \ce{H2O} molecules in a $15\times15$ cell for cluster, 6 \ce{H2O} in a $3\times3$ cell for monolayer, and 10 \ce{H2O} molecules in a $3\times3$ cell for bilayer. After defining the initial positions of the water molecules, the entire system was initialized with momenta corresponding to \qty{10}{\kelvin} using the Maxwell-Boltzmann distribution. Any non-zero rotation or center of mass translation of the system was removed before the MD run. The final structure of each MD simulation was subsequently relaxed using the BFGS algorithm, with a force convergence criterion of \qty{0.01}{\eV\per\angstrom}.

\begin{figure*}
    \centering
    \includegraphics[width=\textwidth, trim = {0 5cm 0 5cm}]
    {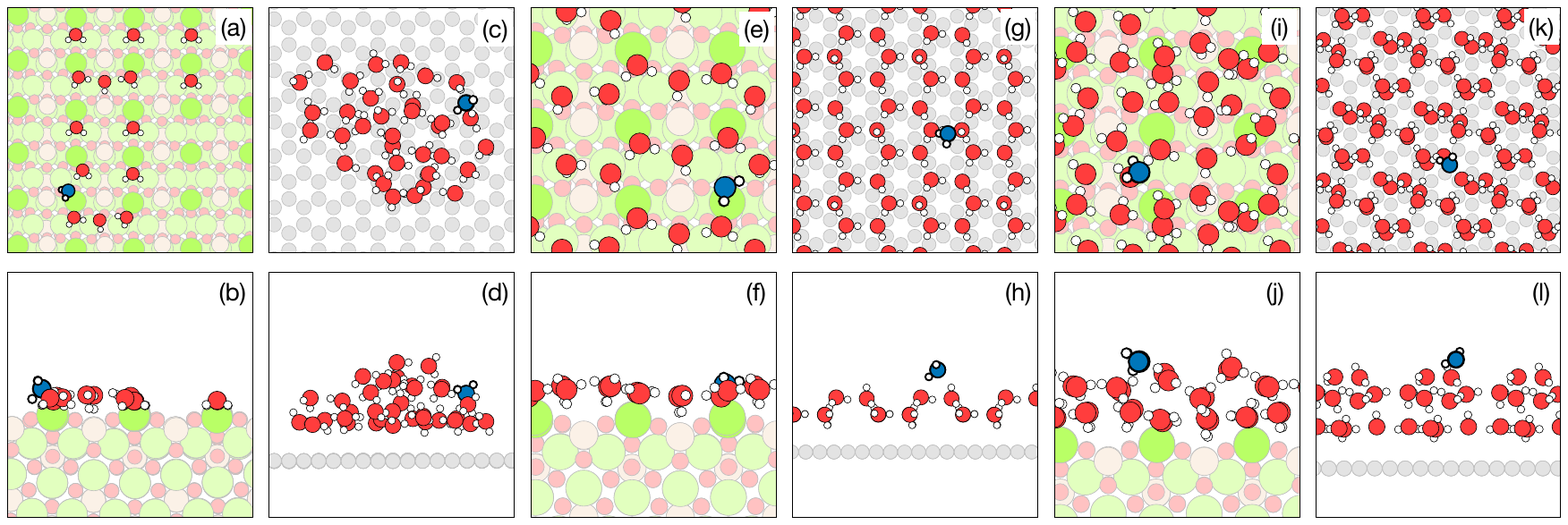}
    \caption{Examples of surfaces generated using global optimization, covering clusters, monolayers and bilayers on the (010) surface facet of forsterite and graphene. Water molecules that are part of the surface are shown with the O atom in red, and a water adsorbate used to probe the BE of an exemplary site is shown with the O atom in blue. Color code for other atoms: H (white), Mg (green, with darker green representing higher height), Si (brown) and C (grey).}
    \label{fig:example structures}
\end{figure*}

Structure generation using GO was done with the GOFEE algorithm. 
For the silicate, the generated surfaces contained 14 \ce{H2O} molecules in a $5\times5$ cell for cluster, 10 \ce{H2O} molecules in a $2\times2$ cell for monolayer and 20 \ce{H2O} molecules in a $2\times2$  cell for bilayer. For the graphene, the generated surfaces contained 35 \ce{H2O} molecules in a $12\times12$ cell for cluster, 6 \ce{H2O} in a $3\times3$ cell for monolayer, and 10 \ce{H2O} molecules in a $3\times3$ cell for bilayer. Examples of the different types of surfaces generated are shown in Figure \ref{fig:example structures} for the case of the GO generation method.

\subsection{Binding energy sampling}

For each combination of structure generation method and type of structure, the adsorbate (\ce{H2O}) was placed on five different structures to obtain more statistically significant BE distributions. For structures generated using GO, the five most stable structures were chosen, while for MD, five structures with different random starting configurations were used. At each surface, the BEs of the various sites exposed was probed one at a time by placing \ce{H2O} onto a given site and relaxing the structure.

Due to the different morphology of the generated surfaces, we used two different methods for placing the adsorbate. For all mono- and bilayer structures and for clusters on silicate, a grid in the $xy$-plane of the cell (parallel to the surface) was constructed with a lateral spacing between grid points of \qty{2.5}{\angstrom} or smaller.

The height of an individual grid point was based on the height of the highest atom of the surface within a \qty{2}{\angstrom} distance to the grid point. The probe water molecule was then placed \qty{2.5}{\angstrom} above the obtained height. For clusters on graphene, a different way of placing the adsorbate was needed as the structure is compact and rounded as shown in Figure \num{1}(h). For this, a total of 60 adsorbate positions were selected such that each water molecule is placed within a distance of \qty{2}{\angstrom} to \qty{2.5}{\angstrom} to the cluster atoms, while maintaining the minimum distance criteria of \qty{2}{\angstrom} between any two atoms of different molecules.
 
For each grid point representing a probe adsorbate position, four different calculations were run, each comprising a preselected rotation angle of the adsorbate. This is done to allow the probe molecule to explore different minima of the potential energy surface, if present, at each site. All probe adsorbate configurations were relaxed using BFGS with a force convergence criterion of \qty{0.01}{\eV\per\angstrom}. Different initial adsorbate configurations that after relaxation ended up with similar adsorbate geometries, based on whether the coordinates of the adsorbates were within \qty{1}{\angstrom} of each other, were filtered such that only the most stable of these geometries were kept.

Energies of those most stable adsorbate configurations are reported as adsorption energies, i.e. the negative of the BE, where more negative represents more stable adsorption:
\begin{equation}\label{AE}
E_{\rm{ads}} = E_{\ce{H2O}-\rm{surface}} - E_{\ce{H2O}} - E_{\rm{surface}}
\end{equation}
Note that no zero-point vibrational energy (ZPVE) corrections are applied. For \ce{H2O} BEs on amorphous solid water ice, ZPVE-corrected numbers can be obtained by applying a scaling factor of 0.854 to the uncorrected energy, as proposed by Ferrero et al.\cite{Ferrero_2020}

The number of hydrogen bonds of each relaxed adsorbate was calculated using a method also applied in our previous work,\cite{Mikkelsen2025} which defines a spherical sector aligned along the vector pointing from the oxygen of the \ce{H2O} towards its hydrogen. A hydrogen bond is considered to exist if an oxygen atom belonging to another \ce{H2O} or to a surface O is present in the spherical sector. The angles and distances defining the spherical sector were set as in our previous work.\cite{Mikkelsen2025} For the \ce{H2O} molecule in focus, the hydrogen bond is categorized as donor if an H atom in the \ce{H2O} points towards another O and as acceptor if the H atom of another \ce{H2O} points towards the O atom in the \ce{H2O}.

In water-covered silicates, besides hydrogen bonds, interactions between surface Mg and O in the adsorbate also play an important role. This Mg-O bond is considered to be present when the distance between a surface Mg atom and the O atom of the \ce{H2O} molecule is within \qty{2.5}{\angstrom}.


\section{Results and Discussion}





\subsection{Effect of grain surface type}




We begin by investigating the effect of the underlying grain surface type (graphene vs.\ silicate), focusing first on cluster and monolayer ice structures generated using the GO method. For clusters, the structure is highly dependent on the grain surface type, as shown in Figure \ref{fig:example structures}(a--d). On the silicate surface, the dominant interaction is the strong Mg-O bond that can be formed between the protruding Mg atoms and \ce{H2O}. Furthermore, \ce{H2O} can achieve additional stabilization through hydrogen bonding to two surface O atoms. Hence, \ce{H2O} does not form a compact cluster, but spreads over the surface to maximize Mg-O bonding. 

\begin{figure*}
    \centering
    \includegraphics[width=0.8\textwidth]{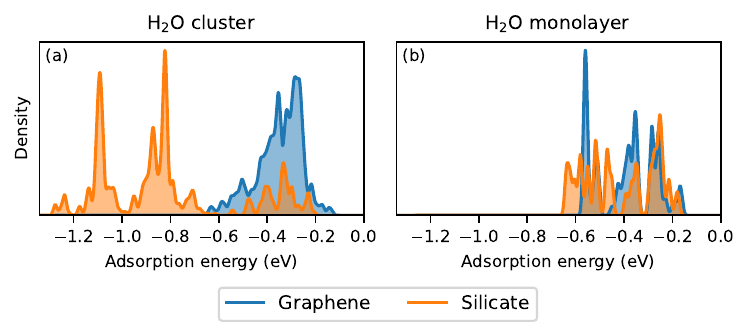}
    \caption{Comparison of BE distributions of \ce{H2O} on graphene and silicate substrates covered by (a) a \ce{H2O} cluster and (b) a \ce{H2O} monolayer. Ice structures were generated with global optimization.}
    \label{fig:graphene_vs_silicate_GO}
\end{figure*}

\begin{figure*}
    \centering
    \includegraphics[width=0.8\textwidth]{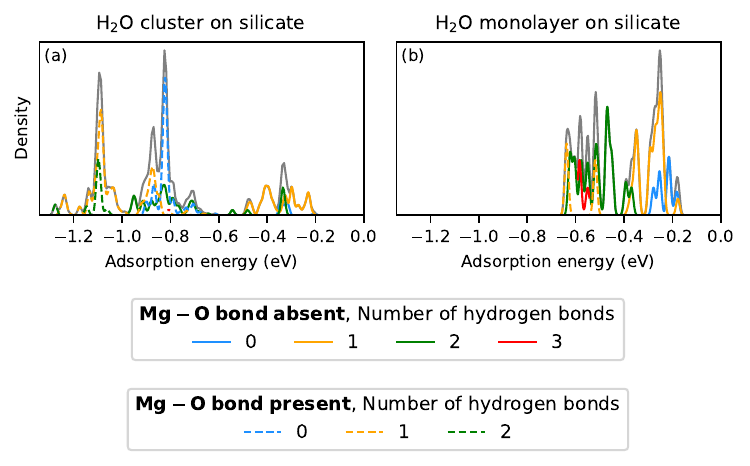}
    \caption{Bonding analysis for (a) cluster and (b) monolayer on silicate for ice structures generated with global optimization. For each \ce{H2O} adsorption configuration, the analysis considers whether a Mg-O bond is present and quantifies the number of hydrogen bonds. The solid gray line is the total density.}
    \label{fig:H-bonding-cluster-monolayer-silicate}
\end{figure*}

\begin{figure*}
    \centering
    \includegraphics[width=\textwidth,trim = {0 2cm 0 2cm}]{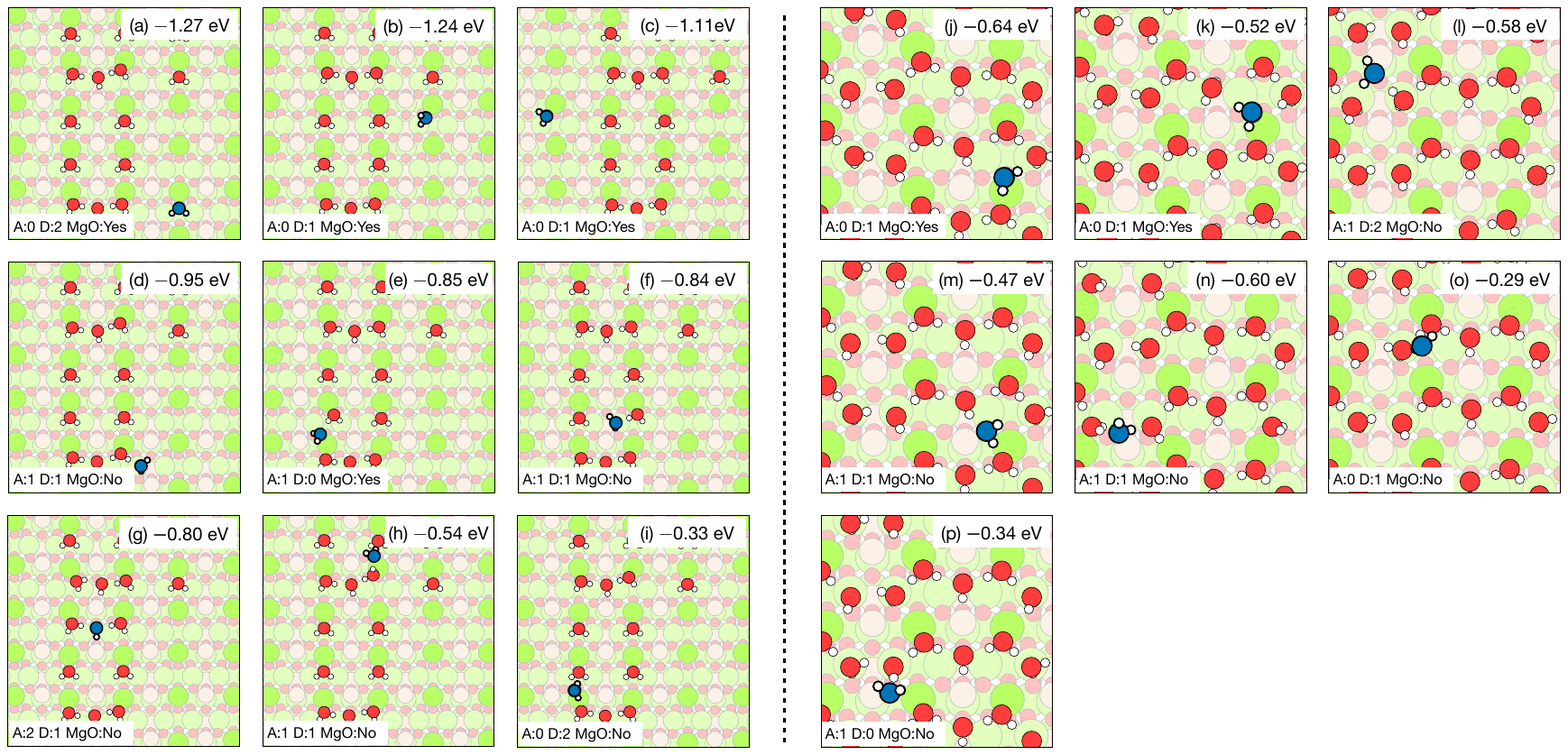}
    \caption{Structures of selected \ce{H2O} adsorption configurations for cluster (left) and monolayer (right). Ice structures were generated with global optimization. The labels indicate the adsorption energy, whether a Mg-O bond is present, and the number of hydrogen bonds of type acceptor (A) and donor (D). The color code is as in Figure \ref{fig:example structures}.}
    \label{fig:fig-H-bonding-cluster-monolayer-silicate}
\end{figure*}

The \ce{H2O} BE distribution on clusters on the silicate surface is very broad compared to clusters on graphene (see Figure \ref{fig:graphene_vs_silicate_GO}(a)). Figure \ref{fig:H-bonding-cluster-monolayer-silicate}(a) presents an analysis of the bonding mechanisms on the silicate surface. As seen, the most stable binding configurations with adsorption energies around $-1.0$~eV to $-1.3$~eV are all characterized by the presence of the Mg-O bond (dashed lines in plot) and one or two hydrogen bonds. Selected structures with adsorption energies in this range are presented in Figure \ref{fig:fig-H-bonding-cluster-monolayer-silicate}(a--c). Clearly, these most stable adsorption motifs are found away from the cluster on the clean silicate surface. The edge of the cluster contains sites with less stable adsorption energies around $-0.8$~eV to $-0.9$~eV (see Figure \ref{fig:fig-H-bonding-cluster-monolayer-silicate}(d--e)) since optimal Mg-O bonding cannot be achieved here. Finally, the top of the cluster contains even less stable sites with adsorption energies in the range $-0.8$~eV to $-0.3$~eV (see Figure \ref{fig:fig-H-bonding-cluster-monolayer-silicate}(f--i)). At these sites there are even more limited possibilities for Mg-O bonding, and the dominant stabilization mechanism comes from hydrogen bonding to cluster \ce{H2O} molecules. Together, these results explain why it is more favorable for incoming \ce{H2O} molecules to spread over the surface, rather than forming a compact cluster.


\begin{figure*}
    \centering
    \includegraphics[width=0.8\textwidth, trim = {0 0cm 0 0cm}]{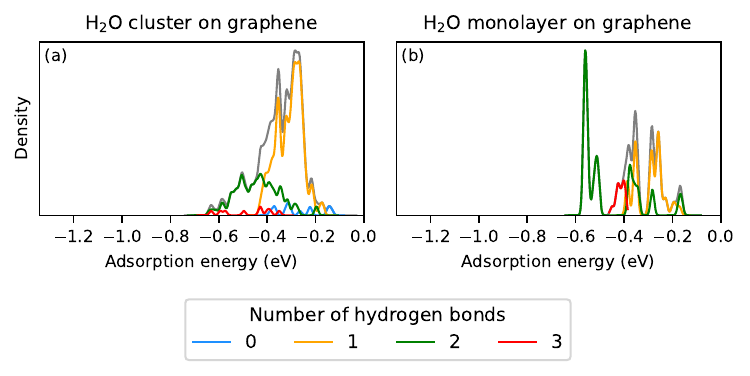}
    \caption{Bonding analysis for (a) cluster and (b) monolayer on graphene for ice structures generated with global optimization. For each \ce{H2O} adsorption configuration, the analysis quantifies the number of hydrogen bonds. The solid gray line is the total density.}
    \label{fig:H-bonding-cluster-monolayer-graphene}
\end{figure*}

\begin{figure*}
    \centering
    \includegraphics[width=\textwidth,trim = {0 3cm 0 2cm}]{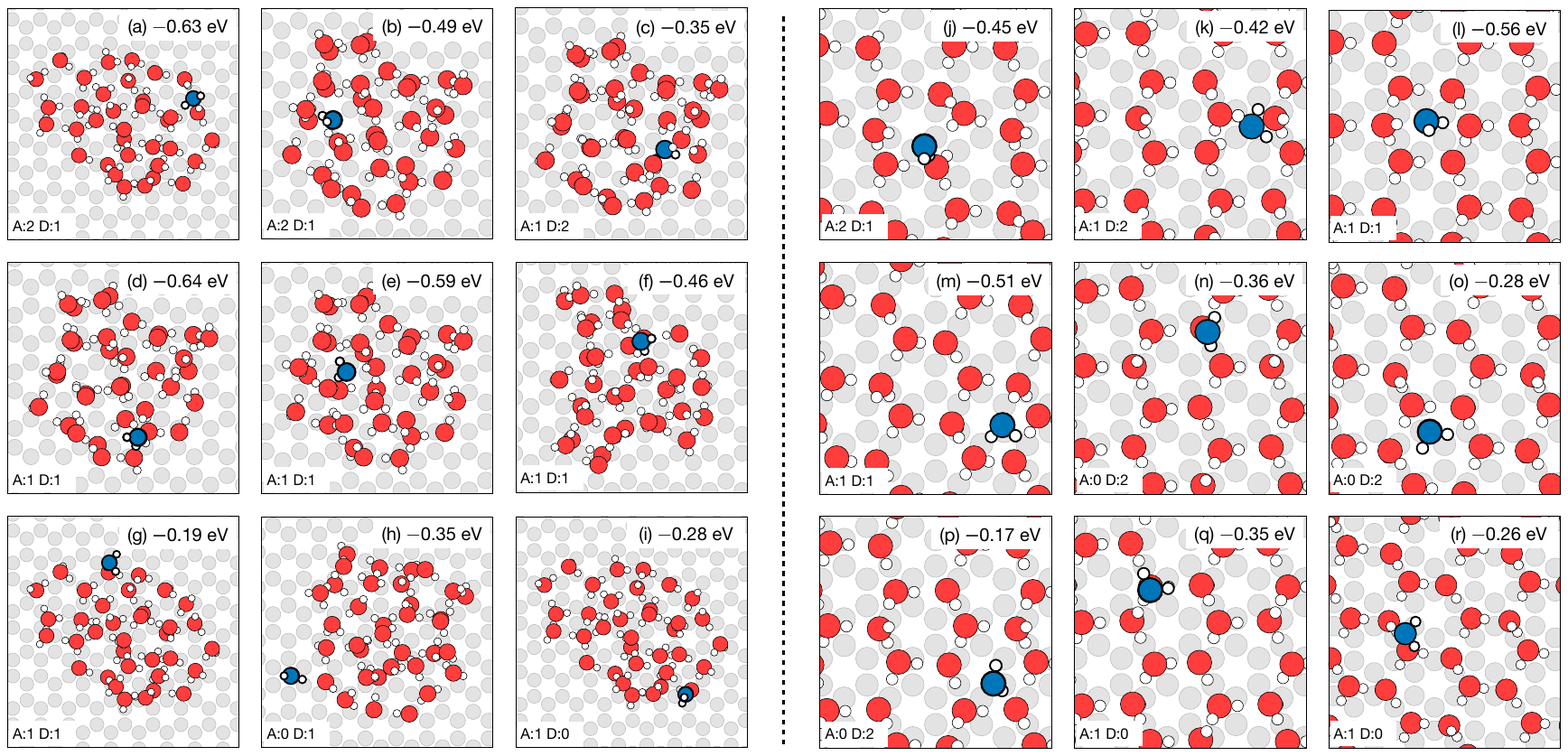}
    \caption{Structures of selected \ce{H2O} adsorption configurations for cluster (left) and monolayer (right). Ice structures were generated with global optimization. The labels indicate the adsorption energy and the number of hydrogen bonds of type acceptor (A) and donor (D). The color code is as in Figure \ref{fig:example structures}.}
    \label{fig:fig-H-bonding-cluster-monolayer-graphene}
\end{figure*}

On the graphene surface, a compact cluster is formed since hydrogen bonding between \ce{H2O} molecules is the dominant interaction (see Figure \ref{fig:example structures}(c)). While the adsorption energy of a single \ce{H2O} on graphene is only about $-0.08$~eV with the trained PaiNN model (see Table S1 in the Supplementary Information), adsorption on top of the cluster is much more favorable with adsorption energies ranging from about $-0.2$~eV to $-0.6$~eV (Figure \ref{fig:graphene_vs_silicate_GO}(a)). There is a rough correlation between the number of hydrogen bonds formed at a given adsorption site and the adsorption energy (see Figure \ref{fig:H-bonding-cluster-monolayer-graphene}(a)). The strongest adsorption energies of about $-0.35$~eV to $-0.6$~eV are found at sites that allow for two or three hydrogen bonds. Selected structures with three hydrogen bonds are shown in Figure \ref{fig:fig-H-bonding-cluster-monolayer-graphene}(a--c) and structures with two hydrogen bonds are shown in Figure \ref{fig:fig-H-bonding-cluster-monolayer-graphene}(d--g). In addition to the number of hydrogen bonds, other neighboring \ce{H2O} molecules and the position on the cluster also play a role. For example, a central position on the cluster with many neighboring \ce{H2O} molecules (Figure \ref{fig:fig-H-bonding-cluster-monolayer-graphene}(d)) gives much stronger adsorption than a position at the edge of the cluster with few neighboring \ce{H2O} molecules (Figure \ref{fig:fig-H-bonding-cluster-monolayer-graphene}(g)), even if two hydrogen bonds are detected for both structures. Finally, adsorption sites near the cluster edge with few hydrogen bonds and neighboring \ce{H2O} molecules results in the least stable adsorption with energies in the range $-0.2$~eV to $-0.35$~eV (see Figure \ref{fig:fig-H-bonding-cluster-monolayer-graphene}(g--i)).

When a complete \ce{H2O} monolayer is formed, the situation changes drastically and the effect of the underlying grain surface type is much less pronounced. Adsorption on the silicate surface is still more favorable with adsorption energies down to $-0.64$~eV, but this is now only slightly lower than for the graphene surface where energies go down to $-0.56$~eV (see Figure \ref{fig:graphene_vs_silicate_GO}(b)). These results demonstrate that, from the monolayer onward, adsorption is dominated by hydrogen bonding with other \ce{H2O} molecules. The small additional stabilization on the silicate surface originates in part from sites that allow for bonding between the O atom in \ce{H2O} with an underlying Mg atom (see bonding analysis in Figure \ref{fig:H-bonding-cluster-monolayer-silicate}(b)). Examples of such structures are shown in Figure \ref{fig:fig-H-bonding-cluster-monolayer-silicate}(j--k). Also the number of hydrogen bonds that can be formed to \ce{H2O} molecules in the monolayer play an important role. The least stable adsorption sites with energies in the range $-0.2$~eV to $-0.35$~eV are characterized by zero or one formed hydrogen bonds, while more stable sites with energies in the range $-0.35$~eV to $-0.6$~eV are characterized by two or three formed hydrogen bonds (see Figure \ref{fig:H-bonding-cluster-monolayer-silicate}(b)). Examples of such structures are shown in Figure \ref{fig:fig-H-bonding-cluster-monolayer-silicate}(l--p).

The monolayer on graphene represents the case with the overall least favorable \ce{H2O} adsorption sites, comprising energies in the range from about $-0.15$~eV to about $-0.55$~eV. This can be attributed partly to the weak interaction with the underlying grain surface (in contrast to the strongly interacting silicate surface) and to the highly stable crystalline structure of the monolayer. The high stability of the monolayer structure considered here---compared to previously proposed structures in the literature \cite{Zhang2018,Yamada2022}---has been discussed in a previous work.\cite{freddy-fractal-water-exp-2025} The bonding analysis in Figure \ref{fig:H-bonding-cluster-monolayer-graphene}(b) shows that, on the monolayer, the correlation between the number of hydrogen bonds formed and the adsorption energy is weak. A couple of sites allow for forming three hydrogen bonds (see Figure \ref{fig:fig-H-bonding-cluster-monolayer-graphene}(j--k)), but with moderate adsorption energies in the range $-0.4$~eV to $-0.45$~eV. The most stable adsorption site occurs in the center of a \ce{H2O} hexagon (Figure \ref{fig:fig-H-bonding-cluster-monolayer-graphene}(l)) with two formed hydrogen bonds and adsorption energy of $-0.56$~eV. Likely, interactions with other neighboring molecules also assist in the stabilization of this motif, even if the bond lengths and angles are not within ranges where the employed classification method categorizes these interactions as hydrogen bonds. Other sites with one or two hydrogen bonds formed (see Figure \ref{fig:fig-H-bonding-cluster-monolayer-graphene}(m--r)) have adsorption energies in the range of about $-0.15$~eV to $-0.5$~eV.

\subsection{Effect of ice structure}
\begin{figure*}
    \centering
    \includegraphics[width=\textwidth, trim={0 3cm 0 2cm}]{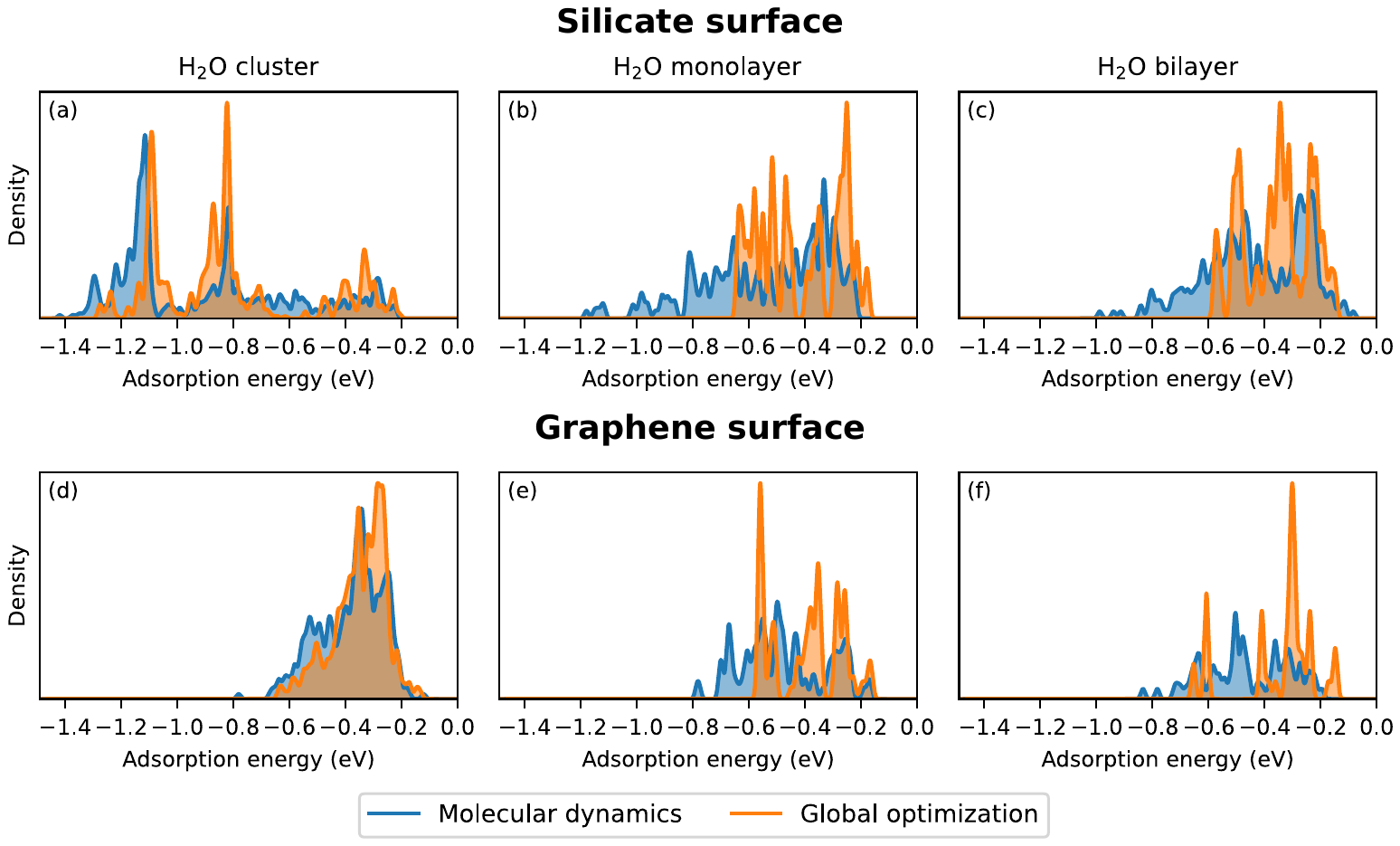 }
    \caption{Comparison of \ce{H2O} BE distributions on ice structures generated using 10~K molecular dynamics and global optimization for cluster, monolayer and bilayer ice structures on silicate (top row) and graphene (bottom row).}
    \label{fig:GO-vs-MD-silicate-graphene}
\end{figure*}

In the previous section we considered stable, crystalline ice structures generated using global optimization. Such structures occur if, during the ice growth or a subsequent warm phase, enough energy has been available for the molecules to explore the potential energy surface. In contrast, ice growth at very low temperatures---such as the conditions found in dense interstellar clouds---generally results in the formation of amorphous ice structures.\cite{Hama2013} To explore the effect of the ice structure (crystalline vs.\ amorphous), we compare in Figure \ref{fig:GO-vs-MD-silicate-graphene} BE distributions on ice clusters, monolayers and bilayers generated using 10~K molecular dynamics simulations and global optimization, respectively, on silicate and graphene surfaces. In all cases, the BE distributions are shifted to lower (more stable) values on the amorphous ices. This effect is particular prominent for the monolayer and bilayer structures. 

\begin{figure*}
    \centering
    \includegraphics[width=\textwidth, trim={2cm 2cm 2cm 2cm}]{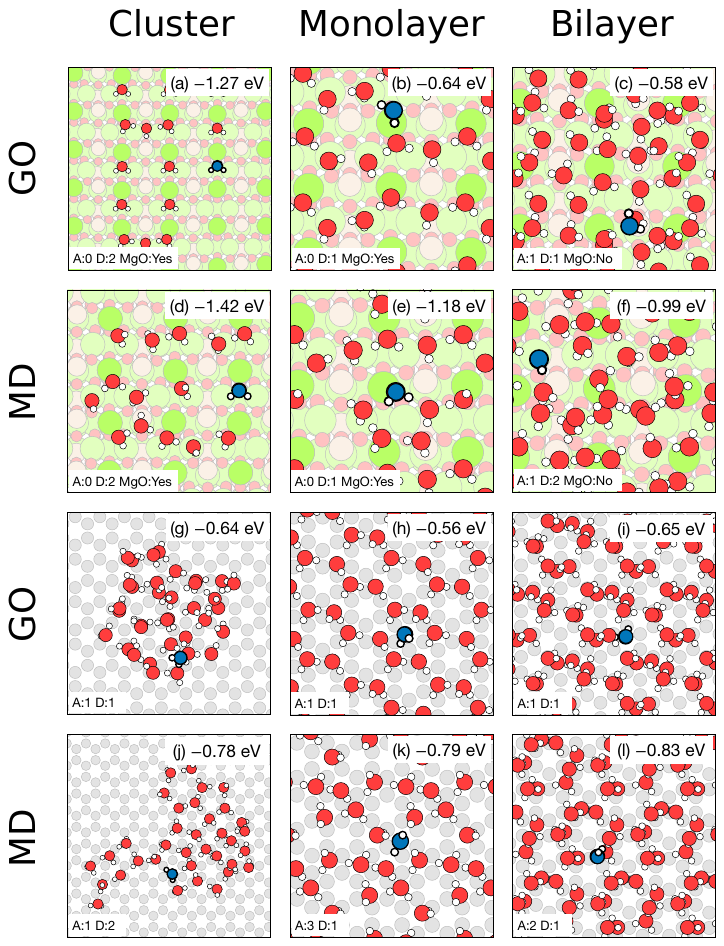}
    \caption{Structures of the most favorable adsorption configurations identified on ice structures generated using low-temperature molecular dynamics and global optimization for cluster, monolayer and bilayer ice structures on silicate (top two rows) and graphene (bottom two rows).}
    \label{fig:GO-vs-MD-silicate-graphene-figures}
\end{figure*}

In Figure \ref{fig:GO-vs-MD-silicate-graphene-figures} we show the most stable adsorption motif found in each case. On the silicate surface, the amorphous monolayer structure clearly covers the grain surface more unevenly, leading to holes where \ce{H2O} molecules can access and bind to Mg atoms in the underlying grain surface. These motifs explain the tail down to $-1.2$~eV in the BE distribution in Figure \ref{fig:GO-vs-MD-silicate-graphene}(b). For the bilayer on silicate, and for both the monolayer and bilayer on graphene, the dominant effect instead appears to be that the rougher amorphous structures contain defect sites that are highly reactive due to the unsaturated hydrogen-bonding capacity of the surrounding \ce{H2O} molecules.

For the clusters, the downward shift in the BE distribution is less pronounced, though still evident. In particular, on the graphene surface it is noteworthy that low-temperature MD growth leads to a porous, fractal morphology with pockets that create especially favorable adsorption sites. The formation of such fractal structures is consistent with recent experimental work on low-temperature water ice nucleation on graphite studied using scanning tunneling microscopy.\cite{freddy-fractal-water-exp-2025}

\section{Conclusions and outlook}

We have trained MLIPs to study \ce{H2O} ice growth and BE distributions on two dust grain models; one representing a case of strong interaction with \ce{H2O} (the forsterite (010) surface) and the other a case of weak interaction with \ce{H2O} (the graphene surface). We show that strong bonding between O in \ce{H2O} and Mg atoms in the forsterite surface gives rise to adsorption energies down to $-1.27$~eV (for an isolated \ce{H2O} molecule) and makes the ice spread over the surface. On graphene, a compact or fractal ice cluster (depending on the growth conditions) dominated by hydrogen bonding between \ce{H2O} molecules is formed instead.

We further assessed the role of the ice structure by generating structures using both global optimization (thus mimicking that enough energy for thermal processing of the ice structure has been available) and through 10~K molecular dynamics simulations (thus mimicking conditions with very little energy present such as dense and cold molecular clouds). We show that ice clusters grown at low temperature have an amorphous and more porous structure with pockets that can bind \ce{H2O} quite strongly, even down to about $-0.8$~eV on the otherwise weakly interacting graphene surface.

Previous literature studies have quantified \ce{H2O} BE distributions using DFT or ONIOM methods on pure, amorphous \ce{H2O} cluster and slab models.\cite{Ferrero_2020,Duflot2021,Bovolenta_2022,Tinacci2023} Adsorption energies (corrected for ZPVE) obtained in these studies range from about $-0.13$~eV to about $-0.64$~eV for the largest cluster model considered.\cite{Tinacci2023} Assuming that the scaling factor of 0.854 proposed by Ferrero et al.\cite{Ferrero_2020} to correct for ZPVE also holds for our structures, the most favorable adsorption sites we identify on the amorphous clusters, mono- and bilayers on graphene have comparable adsorption energies to the most favorable sites identified by Tinacci et al.\cite{Tinacci2023} Based on models from their work, such strong binding sites may contribute to water enrichment of rocky planets inside the classical snowline (i.e., the snowline derived from single-BE astrochemical models) in protoplanetary disks and potentially have implications for the origin of terrestrial water. 

Similarly, the strong binding of \ce{H2O} on silicate grains has previously been discussed in the context of trapping of \ce{H2O} in warm astrophysical environments.\cite{Muralidharan2008,Potapov_2024} Here we show that sites with adsorption energies as low as $-1.3$~eV to $-1.4$~eV (without ZPVE corrections, which have been found to reduce the BE by about $0.12$~eV on forsterite\cite{Molpeceres2018}) are present at low \ce{H2O} coverage. Depending on the ice structure, stabilization through bonding to Mg atoms may still be present at the monolayer coverage, but is mostly absent from bilayer coverage onward.

The BE trends identified here using \ce{H2O} as the probe adsorbate may be transferable to other adsorbates with hydrogen bonding capacity. As the BE impacts diffusion and reaction barriers, the wide range of BEs present at silicate and carbon-based grains in the submonolayer to few-layer ice coverage regime may have important implications for the efficacy of reaction paths towards complex organic molecules. For instance, previous DFT-based works have calculated reaction paths for glycolaldehyde formation through the condensation of formaldehyde \cite{Vinogradoff2024} and for glycine formation through reactions of formaldehyde, carbon monoxide and ammonia over silicate surfaces.\cite{Mates-Torres2026} Substantial modifications to the obtained reaction path energy profiles would be anticipated in the case of submonolayer to few-layer \ce{H2O} ice coverage, as this would allow for hydrogen bonding to reaction intermediates. The ice structures and binding sites identified in the present work may thus find use in future studies targeting diffusion or reactivity.

\section*{Associated content}
	
\subsection*{Supporting Information}
	
Additional benchmarking results and structures and parity plots related to the MLIP training.
	
\subsection*{Data Availability}
Structures are available on Zenodo at \url{https://doi.org/10.5281/zenodo.18242881} under a CC BY 4.0 license.
	
\begin{acknowledgement}
We acknowledge funding from VILLUM FONDEN (grant no.\ 37381) and from the Danish National Research Foundation through the Center of Excellence ‘InterCat’ (grant agreement no.\ DNRF150).
\end{acknowledgement}

\bibliography{bibliography}

\clearpage
	
	

\end{document}